# Automated COVID-19 CT Image Classification using Multi-head Channel Attention in Deep CNN

Abhiroop Chatterjee
*Department of Computer Science and Engineering*
*Jadavpur University*
Kolkata, India
abhiroopchat1998@gmail.com

Susmita Ghosh
*Department of Computer Science and Engineering*
*Jadavpur University*
Kolkata, India
susmitaghoshju@gmail.com

*Abstract—* The rapid spread of COVID-19 has necessitated efficient and accurate diagnostic methods. Computed Tomography (CT) scan images have emerged as a valuable tool for detecting the disease. In this article, we present a novel deep learning approach for automated COVID-19 CT scan classification where a modified Xception model is proposed which incorporates a newly designed channel attention mechanism and weighted global average pooling to enhance feature extraction thereby improving classification accuracy. The channel attention module selectively focuses on informative regions within each channel, enabling the model to learn discriminative features for COVID-19 detection. Experiments on a widely used COVID-19 CT scan dataset demonstrate a very good accuracy of 96.99% and show its superiority to other state-of-the-art techniques. This research can contribute to the ongoing efforts in using artificial intelligence to combat current and future pandemics and can offer promising and timely solutions for efficient medical image analysis tasks.

## I. Introduction

The outbreak of COVID-19 has raised significant challenges for healthcare systems worldwide. Medical imaging, such as CT scans, plays a crucial role in the detection and diagnosis of the disease. In this research, we propose a novel deep learning framework for the automated classification of COVID 19 from SARS-CoV-2 CT scan dataset [1]. The proposed approach leverages the power of combination of a newly designed channel attention mechanism and weighted global average pooling in deep convolutional neural networks (Xception Net) to capture essential features and enhance model's performance.

Classification is one of the important steps in machine learning/pattern recognition/image processing where the model initially learns from a sufficient amount of labeled data and once learnt, the model is expected to be able to generalize well. However, some of the challenges in image classification include imbalanced data, lack of quality training data, selection of set of features to be considered for classification etc. Several researchers worked towards these directions. Various types of artificial neural networks, in isolation [2-4], or in combination with fuzzy logic and/or evolutionary computation methods [5-7] are widely adopted in this regard. Later on deep learning [8-10] gained huge popularity due to its ability to extract complex features from images. The architecture of the proposed model is based on the modified version of the widely used Xception [11] model. On this baseline model, a novel channel attention mechanism is introduced, which focuses on learning relevant features within each channel to improve the model's ability to distinguish COVID-19 positive scans from healthy ones.

The proposed attention module is integrated into the model, enabling it to automatically identify significant patterns within each channel while suppressing less informative regions. By doing so, the model can learn more discriminative representations, leading to accurate and reliable COVID-19 classification.

To evaluate the performance of our proposed approach, experiments were conducted on a popularly used COVID-19 CT scan dataset. Performance comparison with several other deep neural network models is done. It is seen that the proposed framework shows its superiority in terms of accuracy and generalization.

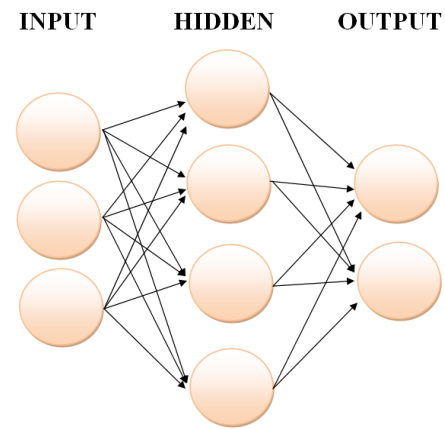

**Fig. 1:** General representation of a neural network.

The present research contributes to the ongoing efforts in using artificial intelligence to combat the COVID-19 and also be prepared for any future pandemics. The proposed model demonstrates the potential of leveraging deep learning and attention mechanisms for accurate and efficient COVID-19 CT scan classification.

The rest of the paper is organized as follows: Section 2 provides a review of related literature in the field of image classification using deep learning. Section 3 presents the methodology, including a detailed description of the proposed channel attention block, its integration within the pretrained Xception network architecture. Section 4 discusses the experimental setup mentioning dataset used, evaluation metrics considered and details of parameters taken. Analysis of results has been put in Section 5. Conclusive remarks have been mentioned in Section 6.

## II. RELATED RESEARCH

Apostolopoulos and Mpesiana [12] in their study proposed an automatic COVID-19 detection system using X-ray images. Leveraging transfer learning with convolutional neural networks (CNNs), the framework demonstrates the potential to accurately diagnose COVID-19 cases with an accuracy of 96.78% for X-ray image dataset. Hemdan et al. [13] presented COVIDX-Net, a comprehensive framework utilizing deep learning classifiers for the diagnosis of COVID-19 from X-ray images. They achieved COVID-19 classification with F1-scores of 0.89 and 0.91 for normal and COVID-19, respectively. In another work, Vruddhi Shah et al. [14] present a study focused on early diagnosis of COVID-19 using CT scan images and deep learning techniques. They developed a novel model, CTnet-10, achieving an accuracy of 82.1% in differentiating between COVID-19 and non-COVID-19 CT scans. Additionally, they tested various deep learning models and found that VGG-19 demonstrated superior performance with an accuracy of 94.52%. In another work Harsh et al. [15] proposes a deep transfer learning algorithm that uses chest X-ray and CT-scan images for fast and accurate detection of COVID-19 cases. It creates relation between COVID-19 and pneumonia patients based on radiology image patterns, aided by Grad-CAM-based color visualization for clear interpretation. The study proposed by Jaiswal et al. [16] used DenseNet201-based deep transfer learning (DTL) model to classify patients as COVID-19 positive or negative. By utilizing the model's own learned weights on the ImageNet dataset along with a convolutional neural structure, features are extracted from the CT images yielding an accuracy of 96.25%. Mishra et al. [17] explored the use of Deep CNN-based approaches to detect COVID-19 from chest CT images and proposed a decision fusion method, achieving an accuracy of 86%.

## III. METHODOLOGY

In this work transfer learning is used on a popular network like the Xception, freezing it by a certain amount and embedding a novel channel attention mechanism into it. The position of embedding is chosen carefully and placed just after the last layer of the Xception network and just before global average pooling.

The following subsections concentrate on the newly incorporated channel attention mechanism followed by a brief description of the Xception network. Since pretrained Xception network is used, a brief discussion on transfer learning technique is also mentioned.

### (A) Multi-head Channel Attention

The proposed attention block (Fig. 2) implements a novel multi-head channel attention mechanism, for medical image classification tasks to enhance feature extraction and improve classification performance. It takes an input feature tensor, which represents the high-level features extracted by convolutional layers in a neural network. Then it applies a weighted global average pooling operation to convert the feature map into a single-channel representation while preserving the original channel dimension. The essence of the channel attention mechanism lies in the attention heads, each designed to learn distinct channel weights. For each head, the block comprises two dense layers. The first one reduces the number of channels through the use of a parameter, chosen experimentally. ReLU activation function is used here. The second dense layer learns the importance of each channel using the sigmoid activation function. The complete process allows the model to focus on relevant information and suppress less important features, thereby improving the model's ability to distinguish between different classes.

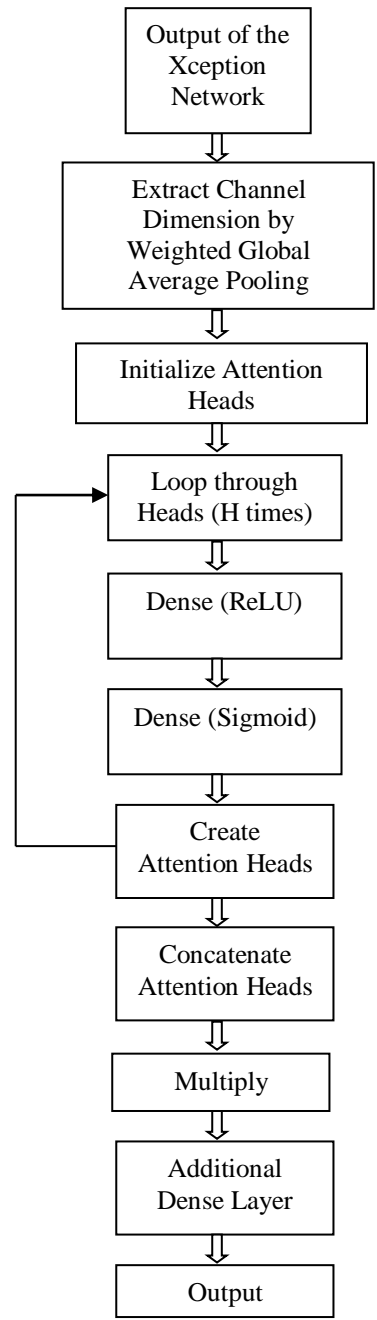

**Fig 2:** Proposed channel attention mechanism.

To ensure diversity and adaptability, the function uses multiple attention heads. The total heads (H) is initialized to 16. Each head learns a unique set of channel weights, introducing a variety of perspectives in the feature selection process. These attention weights and the original weighted global average pooling results are then combined through element-wise multiplication to produce the final output feature, highlighting informative channels while discarding irrelevant ones. The aggregation of attention weights and the original average pooling results is followed by a reshaping operation, ensuring that the feature map aligns with the subsequent dense layer's input shape. An additional dense layer is introduced to further process the output feature, enabling the model to capture higher-level representations and complex relationships within the data.

Overall, the multi-head channel attention mechanism aims to enhance the discriminative power of the model, providing it with the capability to focus on crucial features within each channel while considering multiple perspectives. This technique contributes to improving the performance of the model, making it more adaptive and effective in solving complex tasks, such as COVID-19 CT scan image classification.

### (B) Architecture of the Xception Network

The Xception architecture [11] has proven to be highly effective in various computer vision tasks, such as image classification, object detection, and semantic segmentation. Its efficient design has made it popular in resource-constrained scenarios and has contributed to advancements in deep learning research for image analysis. In the present work, we have used Xception network due of the said advantages over other popular deep neural nets.

The main idea behind Xception network is to use depthwise separable convolutions to reduce the computational complexity of traditional convolutions. In a regular convolution, each filter operates on the entire input volume, leading to a large number of computations. Whereas, depthwise separable convolutions split the convolution into two separate steps: depthwise convolutions and pointwise convolutions.

Here is a brief description of the two main components of Xception model:

1. Depthwise Convolution: In this step, each channel of the input is convolved separately with its corresponding filter, creating a set of intermediate feature maps.
2. Pointwise Convolution: The pointwise convolution follows the depthwise convolution and performs a 1x1 convolution across all the intermediate feature maps. This process helps to combine and aggregate information from different channels efficiently.

The Xception architecture replaces the standard Inception modules used in the original Inception architecture with these depthwise separable convolutions, thereby significantly reducing the number of parameters and computations while maintaining or even improving performance.

### (C) Transfer Learning

In the proposed work a pretrained Xception net, trained on the ImageNet dataset [18], has been used for image classification. The pretrained model is used as a feature extractor for a new custom classification task, rather than training a new model from scratch, which can be time-consuming and computationally expensive.

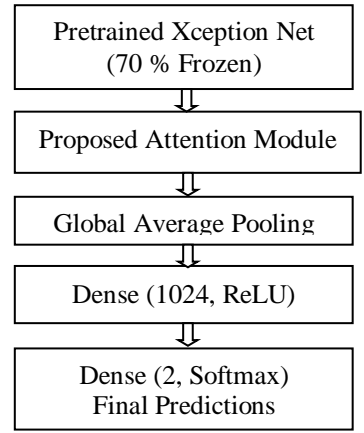

**Fig**. 3: Proposed transfer learning framework.

We start by deploying the Xception model with pretrained weights on ImageNet. The final fully connected layers of the original Xception model was removed, as we want to replace them with our custom classification layers (Fig. 3).

To make the transfer learning process more efficient, we have decided to freeze a certain percentage of the base model's layers. In this case after various experimentations, 70% of the layers were frozen (non-trainable). By freezing these layers, we prevent their weights from being updated during training, which helps to retain the knowledge learned from the ImageNet dataset while focusing on fine-tuning the remaining layers for our specific task.

To build the final classification model, we add our custom layers on top of the frozen Xception base. First, the output of the base model is passed through the proposed channel attention layer, which presumably enhances the feature representations. The multi-head channel attention mechanism is embedded immediately after the output of the base Xception model and before the Global Average Pooling 2D layer. This placement is strategically chosen based on the design of the network and the purpose of using attention in the context of transfer learning. A Global Average Pooling 2D layer is used to reduce the spatial dimensions of the data.

Continuing with the custom classification layers, a dense layer with 1024 neurons and a ReLU activation function is used to learn more complex patterns from images. Finally, another dense layer with softmax activation function is utilized to obtain the probabilities of each of the classes.

Needless to mention that the number of neurons in the last dense layer is determined by the number of classes to be classified.

## IV. EXPERIMENTAL SETUP

As stated, experiment with the proposed framework is conducted using the SARS-CoV-2 CT scan dataset. Details of this experimental setup have been described below.

**(A) Dataset Used:**

As mentioned earlier, SARS-CoV-2 CT scan dataset [1] is used. It consists of 1252 images of COVID-19 and 1230 images of non-COVID cases (Table 1). Sample images from both the classes have been shown in Fig. 4.

**(B) Image Preprocessing:**

Each image is resized to a dimension of 224x224 pixels and normalized by dividing the pixel values by 255.

**Table 1:** Images taken from SARS-CoV-2 CT dataset [1].

| Types of Classes | Number of Images |
|---|---|
| COVID-19 | 1252 |
| NON COVID-19 | 1230 |
| **Total Images** | **2482** |

**(C) Performance Metrics Considered:**

The models' performance is evaluated using metrics such as accuracy, loss values. We have also considered other performance metrics e.g., precision, recall, F1-score, Top-1% error for evaluation purposes. Confusion matrix has also been considered.

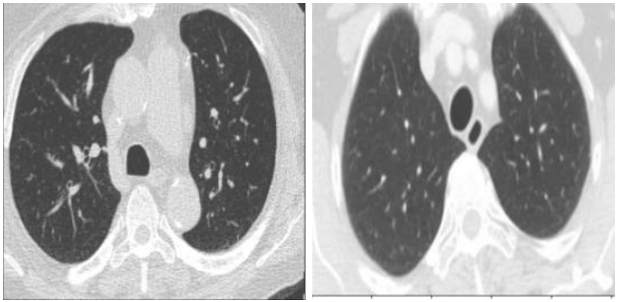

**Fig. 4:** Images taken from SARS-CoV-2 CT scan dataset (a) Covid-19, (b) Non Covid-19.

**(D) Parameters Taken:**

Table 2 shows the parameters and their corresponding values used for our experimentation.

**(E) Model Training:**

The dataset is split into training and test sets. The split is performed with a test size of 20% and stratified sampling to maintain class balance. During training, the models' weights are updated using back-propagation and gradient descent to minimize the loss function.

**TABLE 2:** Experimental setup

| Parameters | Values |
|---|---|
| Learning Rate | 0.0001 |
| Batch Size | 16 |
| Max Epochs | 50 |
| Optimizer | Adam |
| Loss Function | Categorical Cross-entropy |

## V. ANALYSIS OF RESULTS

To evaluate the effectiveness of the incorporation of the proposed multi-head channel attention block in transfer learnt Xception network, experiment was conducted on SARS-CoV-2 dataset. A total of 20 simulations have been performed and the average values (over 20 simulations) obtained for each of the metrics are depicted into Table. 3. From the table it is noticed that the results are promising in nature in terms of various performance indices, yielding 96.99% accuracy for the transfer learnt modified Xception model. Experimentation was done on NVIDIA A100 tensor core GPU.

**TABLE 3:** Results obtained using SARS-CoV-2 CT scan dataset for proposed framework.

| Metrics | Average Score |
|---|---|
| Precision | 0.96995 |
| Recall | 0.97000 |
| F1-score | 0.96997 |
| Accuracy (%) | 96.99 |
| Top-1 error (%) | 3.01 |

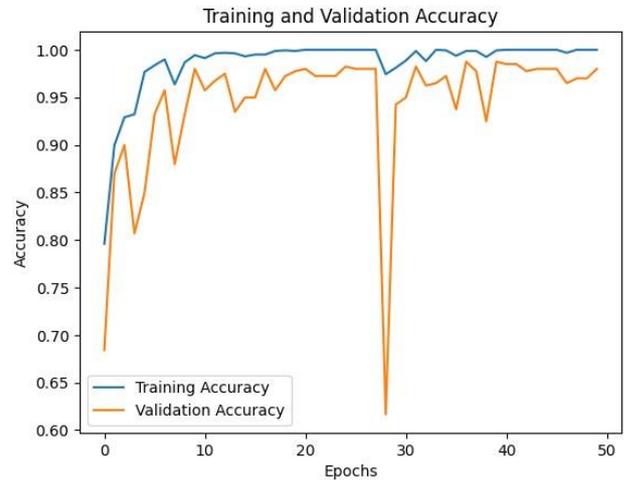

**Fig 5:** Variation of training and validation accuracy with epochs of the fine-tuned Xception model embedded with proposed attention module.

Variations of training & validation accuracy and training & validation loss over epochs for the proposed framework are shown in Figs. 5 and 6, respectively. In these figures the results for best simulation (out of 20) have been depicted. From Fig. 5 it is seen that both training and validation accuracy increase over epochs. This suggests that the model is generalizing the data well and can make accurate predictions. As training progresses, the validation and training accuracies continue to steady down. Likewise, as expected, loss values are also getting stabilized over epochs (Fig. 6).

To compare the performances of the baseline Xception model with that of the modified Xception net (embedded with channel attention) variations of training accuracy and validation accuracy (best over 20 simulations) for both the cases over epochs are displayed in Figs. 7 and 8, respectively. It is seen that both models start to steady down after 10 epochs which indicates that a significant learning progress is achieved early in the training process. A similar trend in the plot is observed for all 20 runs which show the credibility to the robustness of the model.

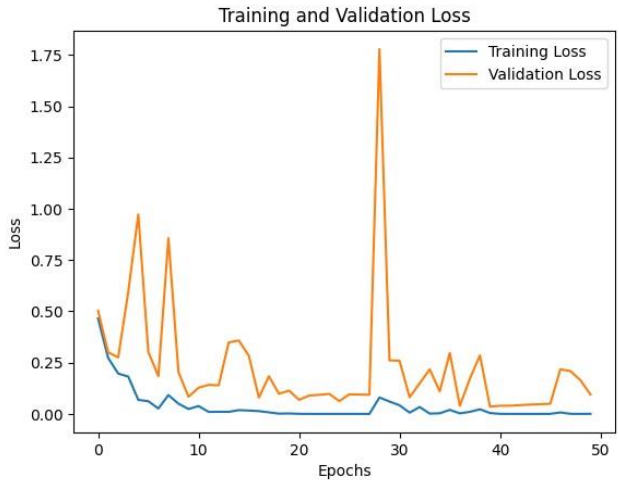

**Fig. 6:** Variation of training loss and validation loss with epochs of the fine-tuned Xception model embedded with proposed attention module.

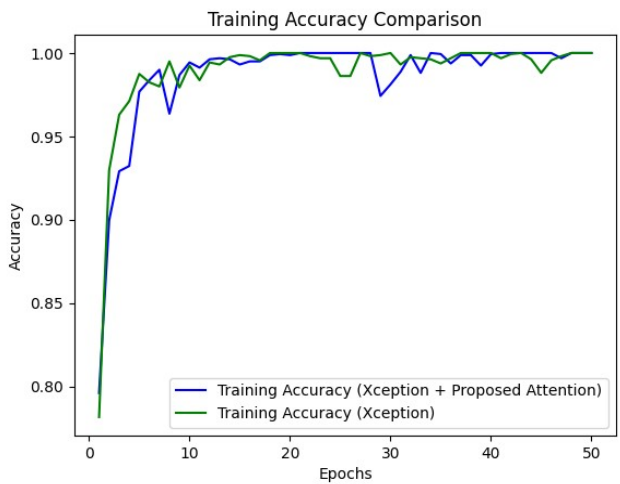

**Fig 7:** Variations of training accuracy with epochs of the fine-tuned Xception model and fine-tuned Xception model embedded with proposed attention module.

From Fig. 8, the validation curves reveal interesting insights between the Xception net and the proposed framework. Despite fluctuations observed in both the curves, the proposed framework consistently outperforms Xception net across almost every epoch. This suggests that the proposed model demonstrates superior generalization capabilities, mitigating overfitting and adapting well to unseen data. The robustness of the proposed framework's performance is evident, leading to reduced generalization gaps and more stable convergence. The observed advantages may be attributed to the embedded attention mechanism along with the model's architectural choices, potential learning rate selection, and effective hyperparameter tuning. Overall, the validation curve analysis indicates the potential of the proposed framework as a more reliable solution for the given image classification task compared to Xception net.

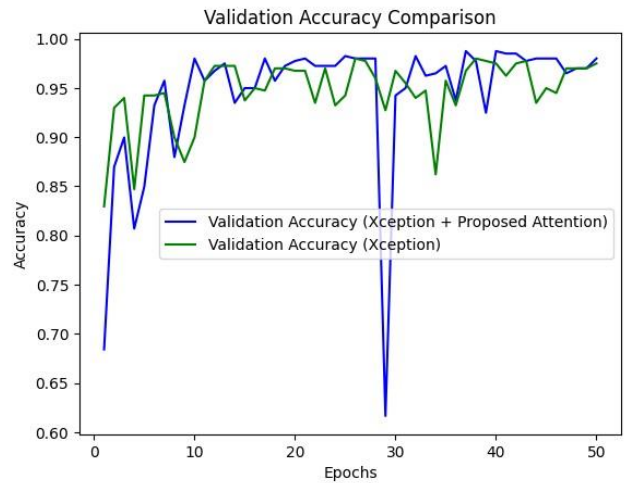

**Fig. 8:** Variations of validation accuracy with epochs of the fine-tuned Xception model and fine-tuned Xception model embedded with proposed attention module.

Based on the values in the confusion matrix, as shown in Fig. 9, the proposed model exhibits impressive performance for COVID-19 detection. With a false positive rate of only 1.8%, the model incurs very few errors in incorrectly classifying healthy individuals as COVID-19 positive. This low false positive rate is crucial as it ensures that the model does not cause unnecessary panic or misdiagnose healthy individuals.

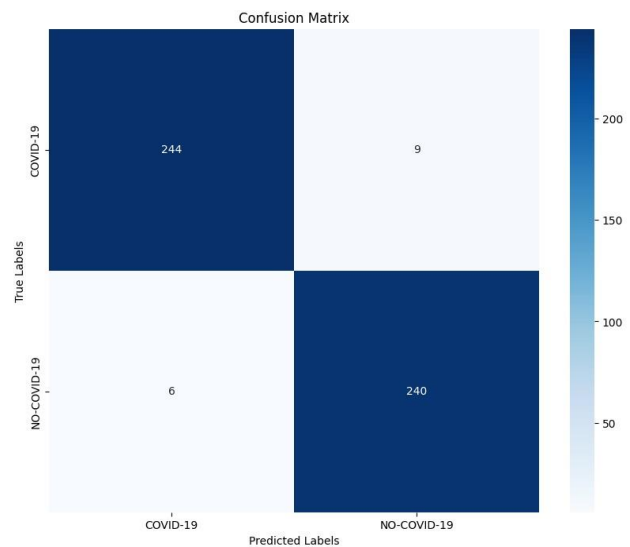

**Fig. 9:** Confusion matrix for COVID-19 detection for the proposed framework.

On top of it, the false negative rate of 1.2% indicates that the proposed model effectively captures the majority of COVID-19 positive cases, minimizing the risk of missing true positive cases and spreading the virus.

As stated, performance of the proposed model has been compared with other deep neural nets and the corresponding accuracy values are shown in Table 4. This table confirms the superiority of the proposed framework as compared to three other fine-tuned networks (AlexNet, VGG-16,VGG-19) for COVID-19 detection. The proposed framework also outperforms baseline Xception net, DTL [14] and the method proposed by Mishra et al. [15]. Through rigorous experiments and comparisons with state-of-the-art methods, the framework achieves higher classification accuracy of 96.99%, underscoring its potential as an advanced and effective solution for accurate COVID-19 diagnosis.

**TABLE 4:** Classification performance of the proposed and the state-of-the-art models for the SARS-CoV-2 CT scan dataset.

| Methods | Accuracy (%) |
|---|---|
| AlexNet (Pretrained on ImageNet) | 93.71 |
| VGG-16 (Pretrained on ImageNet) | 94.62 |
| VGG-19 (Pretrained on ImageNet) | 93.56 |
| Jaiswal et al. [16] | 96.25 |
| Mishra et al. [17] | 88.34 |
| Xception (70% frozen) | 95.39 |
| Proposed Xception embedding multi-head channel attention | **96.99** |

## V. CONCLUSION

In this article, we presented a novel approach to enhance image classification performance using a combination of transfer learning and a multi-head channel attention mechanism. Our methodology leverages the potential of powerful Xception model pretrained with ImageNet dataset. By transferring knowledge from the Xception model, training time and computational resources are reduced, enabling us to focus on fine-tuning the model through channel attention. This channel attention module is placed immediately after the output of the base Xception model. The attention mechanism dynamically highlighted relevant features in the intermediate feature maps, providing the model with the ability to focus on the most informative regions for image classification.

The results of this research indicate that transfer learning, combined with the channel attention mechanism, is a powerful and efficient approach to address image classification challenges. Moreover, the computational efficiency of the proposed approach makes it well-suited for deployment on resource-constrained systems. As future work, we suggest exploring the application of this methodology on other computer vision tasks, such as object detection and semantic segmentation. Investigating the impact of different attention mechanisms or modifying the architecture to accommodate attention at different levels may further improve performance.

## Acknowledgement

A part of this work has been supported by the IDEAS - Institute of Data Engineering, Analytics and Science Foundation, The Technology Innovation Hub at the Indian Statistical Institute, Kolkata through sanctioning a Project No. /ISI/TIH/2022/55/ dtd. September 13, 2022.